\documentclass[12pt]{article}

\usepackage{amssymb,amsfonts,amsmath,bm,color,cite,ulem}
\usepackage{graphicx}

\begin{document}




\vskip 2cm

\begin{center}
{\Large\bf Oscillating 4-Polytopal Universe in Regge Calculus}
\end{center}
\vspace*{1cm}
\begin{center}{\sc Ren Tsuda$^{1}$ and Takanori Fujiwara$^2$}
\end{center}
\vspace*{0.2cm}
\begin{center}
{\it $^1$Student Support Center, Chiba Institute of Technology, Narashino 275-0023, Japan} \\
{\it $^2$Department of Physics, Ibaraki University, Mito 310-8512, Japan}
\end{center}

\vfill




\begin{abstract} 
The discretized closed Friedmann--Lema\^itre--Robertson--Walker (FLRW) universe with 
positive cosmological constant is investigated by Regge calculus. 
According to the Collins--Williams formalism, a hyperspherical Cauchy surface is replaced 
with regular 4-polytopes. Numerical solutions to the Regge equations approximate well 
to the continuum solution during the era of small edge length. Unlike the expanding 
polyhedral universe in three dimensions, the 4-polytopal universes repeat expansions 
and contractions. To go beyond the approximation using regular 4-polytopes we introduce 
pseudo-regular 4-polytopes by averaging the dihedral angles of the tessellated regular 
600-cell. The degree of precision of the tessellation is called the frequency. Regge 
equations for the pseudo-regular 4-polytope have simple and unique expressions for 
any frequency. In the infinite frequency limit, the pseudo-regular 4-polytope model 
approaches the continuum FLRW universe. 
\end{abstract}

\newpage

\pagestyle{plain}




\section{Introduction}

\label{sec:intro}
\setcounter{equation}{0}

Regge calculus was proposed in 1961 to formulate Einstein's 
general relativity on piecewise linear manifolds 
\cite{Regge:1961aa, MTW}. It is a coordinate-free discrete 
formulation of gravity, providing a framework in both classical and 
quantum studies of gravity \cite{BOW:2018aa}. Since Regge calculus is a highly 
abstract and abstruse theoretical formalism based on simplicial 
decomposition of space-time, further theoretical studies from various 
viewpoints are to be welcomed. In particular, exact results such as the 
Friedmann--Lema\^itre--Robertson--Walker (FLRW) universe and Schwarzschild 
space-time in the continuum theory are considered to play the role of 
a touchstone in Regge calculus. Thorough investigations on these systems 
are indispensable in making Regge calculus practical for use. 

Regge calculus has been applied to the four-dimensional closed 
FLRW universe by Collins and 
Williams \cite{CW:1973aa}. They considered 
regular polytopes (4-polytopes or polychora) as the Cauchy surfaces of 
the discrete FLRW universe 
and used, instead of simplices, truncated world-tubes evolving from one 
Cauchy surface to the next as the building blocks of piecewise linear 
space-time. Their method, called the Collins--Williams (CW) formalism, is 
based on the $3+1$ decomposition of space-time similar to the 
well-known Arnowitt--Deser--Misner (ADM) formalism \cite{AD:1959aa,ADM:1959aa,Brewin:1987aa}. 
Recently, Liu and Williams have extensively studied the discrete FLRW 
universe \cite{LW:2015aa,LW:2015ab,LW:2015ac}. They found that a universe with 
regular 4-polytopes such as the Cauchy surfaces can reproduce the continuum 
FLRW universe to a certain degree of precision. Their solutions agree 
well with the continuum when the size of the universe is small, whereas 
the deviations from the exact results become large for a large universe. 

In a previous paper \cite{TF:2016aa}, we investigated 
the three-dimensional closed FLRW universe with positive cosmological 
constant by the CW formalism. The main interest there 
was to elucidate how Regge calculus reproduces the FLRW universe in 
the continuum limit. In three dimensions, a spherical Cauchy surface 
is replaced with polyhedra. We described the five types of regular 
polyhedra on an equal footing by using Schl\"afli symbols \cite{Coxeter}. 
The polyhedral universe behaves as the analytic solution of the continuum 
theory when the size of the universe is small, while it expands to 
infinity in a finite time. We further proposed a geodesic dome model 
to go beyond the regular polyhedra. The Regge calculus for the 
geodesic domes, however, becomes more and more complicated as we 
better and better approximate the sphere. To avoid the cumbersome 
tasks in carrying out Regge calculus for the geodesic dome models
we introduced pseudo-regular polyhedra characterized by fractional Schl\"afli symbols. 
Regge equations for the pseudo-regular polyhedron model turned out 
to approximate the corresponding geodesic dome universe well. 
It is worth investigating whether a similar approach can be extended 
to higher dimensions. 

In this paper we investigate the FLRW universe of four-dimensional Einstein 
gravity with a positive cosmological constant within the framework of the CW 
formalism. We consider all six types of regular 4-polytopes as the 
Cauchy surface in a unified way in terms of the Schl\"afli 
symbol and compare the behaviors of the solutions with the 
analytic result of the continuum theory. We further propose a 
generalization of the Regge equations by introducing pseudo-regular 
polytopes, which makes the numerical analysis much easier than the 
conventional Regge calculus. 

This paper is organized as follows: in the next section we set up the regular 
4-polytopal universe in the CW formalism and introduce Regge action. 
Derivation of the Regge equations is given in Sect. \ref{sec:req}. In the 
continuum time limit the Regge equations are reduced to differential 
equations. Applying the Wick rotation, we arrive at the Regge calculus 
analog of the Friedmann equations, describing the evolution of the 
4-polytopal universe. This is done in Sect. \ref{sec:ctl}. In Sect. 
\ref{sec:nsol} we solve the differential Regge equations numerically and 
compare the scale factors of the 4-polytopal universes with the continuum 
solution. In Sect. \ref{sec:prpt} we consider subdivision of cells of 
the regular 4-polytopes and propose a pseudo-regular 4-polytopal universe 
with a non-integer Schl\"afli symbol that approaches a smooth 
three-dimensional sphere in the continuum limit. Sect. \ref{sec:sum} is 
devoted to summary and discussions. In Appendix \ref{sec:crrdp}, the 
radius of the circumsphere of a regular polytope in any dimensions 
is considered.




\section{Regge action for a regular 4-polytopal universe}

\label{sec:ra}
\setcounter{equation}{0}

\begin{table}[t]
\begin{align*}
\begin{array}{cccc}
\hline
 & k=1 & k = 0 & k = -1 \\
 \hline
\Lambda > 0 & a = \sqrt{\frac{3}{\Lambda}} \cosh \left( \sqrt{\frac{\Lambda}{3}} t \right) & a = \sqrt{\frac{3}{\Lambda}} \exp \left( \sqrt{ \frac{\Lambda}{3} } t \right) & a = \sqrt{ \frac{3}{\Lambda} } \sinh \left( \sqrt{ \frac{\Lambda}{3} } t \right) \\
\Lambda = 0 & \mbox{no solution} & a = \mbox{const.} & a = t \\
\Lambda < 0 & \mbox{no solution} & \mbox{no solution} & a = \sqrt{- \frac{3}{\Lambda}} \sin\left( \sqrt{ - \frac{\Lambda}{3} } t \right) \\
\hline
\end{array}
\end{align*}
\caption{Solutions of the Friedmann equations.}
\label{table:4d_flrw}
\end{table}

We begin with a brief description of the continuum 
FLRW universe. The continuum action is given by 
\begin{align}
  \label{eq:cEHa}
  S=\frac{1}{16\pi}\int d^4x\sqrt{-g}(R-2\Lambda).
\end{align}
In four dimensions, the Einstein equations have an evolving universe as a 
solution for the ansatz
\begin{align}
  \label{eq:FLRWm}
  ds^2=-dt^2+a(t)^2\left[\frac{dr^2}{1-kr^2}+r^2 \left( d \theta^2 + \sin^2 \theta d\varphi^2 \right) \right],  
\end{align}
where $ a \left( t \right) $ is the so-called scale factor in cosmology. It is subject to the Friedmann equations
\begin{align}
  \label{eq:Feq}
  \dot a^2=\frac{\Lambda}{3} a^2-k, \qquad
  \ddot a=\frac{\Lambda}{3} a. 
\end{align}
The curvature parameter $k=1,0,-1$ corresponds to space being spherical, 
Euclidean, or hyperbolic, respectively. The relations between 
the solutions and curvature parameter are summarized in Table \ref{table:4d_flrw} with 
the proviso that the behaviors of the universes are restricted to expanding 
at the beginning for the initial condition $a(0)=\min a(t)$.
Of these, our concern is the spherical universe  
with three-dimensional spheres as the 
Cauchy surfaces. All the time dependence of the universe is in the 
scale factor $a(t)$, which expresses the curvature radius of the Cauchy 
surface. In Regge calculus we will replace the three-dimensional 
sphere with a regular 4-polytope. 

Before entering into details of the 4-polytopal universe,
let us briefly summarize the essence of Regge calculus:
in Regge calculus, an analog of the Einstein--Hilbert action is given by 
the Regge action \cite{Miller:1997aa}
\begin{align}
  \label{eq:ract}
  S_{\rm Regge}=\frac{1}{8\pi}\left(\sum_{i\in\rm \{hinges\}}
    \varepsilon_iA_i-\Lambda\sum_{i\in\rm \{blocks\}} V_i\right),
\end{align}
where $A_i$ is the volume of a hinge, $\varepsilon_i$ the deficit angle 
around the hinge $A_i$, and $V_i$ the volume of a building block of 
the piecewise linear manifold. In four dimensions the hinges are the 
lattice planes, or equivalently the faces of the 4-simplices, and $A_i$ 
is nothing but the face area. Regge's original derivation is concerned 
with a simplicial lattice, so that it describes the gravity as simplicial 
geometry. In fact this formalism can easily be generalized to arbitrary lattice 
geometries. We can fully triangulate the non-simplicial flat 
blocks by adding extra hinges with vanishing deficit angles without 
affecting the Regge action. 

The fundamental variables in Regge calculus 
are the edge lengths $l_i$. Varying the Regge action with respect to $l_i$, 
we obtain the Regge equations 
\begin{align}
  \label{eq:regeq}
  \sum_{i\in \rm \{hinges\}}\varepsilon_i\frac{\partial A_i}{\partial l_j}
  -\Lambda\sum_{i\in \rm \{ blocks \}}\frac{\partial V_i}{\partial l_j}=0.
\end{align}
Note that there is no need to carry out the variation of the deficit angles 
owing to the Schl\"afli identity \cite{Schlafli:1858aa, HHKL:2015aa}
\begin{align}
\sum_{i\in\rm\{hinges\}}A_i\frac{\partial\varepsilon_i}{\partial l_j}=0.
\end{align}

\begin{table}[t]
  \centering
  \begin{tabular}{ccccccc}\hline 
    & 5-cell & 8-cell & 16-cell & 24-cell & 120-cell & 600-cell \\ \hline
    $N_3$ & 5 & 8 & 16 & 24 & 120 & 600 \\
    $N_2$ & 10 & 24 & 32 & 96 & 720 & 1200 \\
    $N_1$ & 10 & 32 & 24 & 96 & 1200 & 720 \\ 
    $N_0$ & 5 & 16 & 8 & 24 & 600 & 120 \\ 
    $\{p,q,r\}$ & $~~\{3,3,3\}~~$ & $~~\{4,3,3\}~~$ & $~~\{3,3,4\}~~$ & $~~\{3,4,3\}~~$ & $~~\{5,3,3\}~~$ & $~~\{3,3,5\}~~$ \\ \hline
  \end{tabular}
  \caption{The six regular polytopes in four dimensions.}
  \label{tab:rpt}
\end{table}

We now turn to 4-polytopal universes.
Following the CW formalism, we replace the hyperspherical Cauchy surface by 
a regular 4-polytope. It would be helpful to begin with a description 
of regular 4-polytopes \cite{Coxeter}. As regular polyhedra in three dimensions, a regular 
4-polytope can be obtained by gluing three-dimensional cells of congruent 
regular polyhedra. Any regular 4-polytope can be specified by the Schl\"afli 
symbol $\{p,q,r\}$, where $\{p,q\}$ stands for the Schl\"afli symbol of a 
cell and $r$ the number of cells having an edge of a cell in common. 
It is known that there are only six types of regular 4-polytopes: 5-cell, 
8-cell, 16-cell, 24-cell, 120-cell, and 600-cell, as listed in 
Table \ref{tab:rpt}. 
Incidentally, this can 
be extended inductively to polytopes in arbitrary dimensions. 
In general, a $D$-polytope can be denoted by a set of $D-1$ 
parameters $\{p_2,\cdots,p_D\}$. 

Let us denote the numbers of vertices, edges, faces, and cells of a 
regular 4-polytope $\{p,q,r\}$ by $N_0$, $N_1$, $N_2$, and 
$N_3$, respectively. They satisfy 
$n_2 \left( q , r \right) N_0= n_0 \left( p , q \right) N_3$, 
$rN_1=n_1 \left( p , q \right) N_3$, 
and 
$2N_2=n_2 \left( p , q \right) N_3$, where $n_0 \left( p , q \right)$, $n_1 \left( p , q \right) $, and $n_2 \left( p , q \right) $ are the numbers 
of vertices, edges, and faces of a regular polyhedron $\{p,q\}$, respectively.  
These completely determine the ratios $N_{0,1,2}/N_3$ as 
\begin{align}
  \label{eq:N0_rpt}
  \frac{N_0}{N_3}
  &=\frac{n_0 \left( p , q \right)}{n_2 \left( q , r \right)}
  =\frac{p(2q+2r-qr)}{r(2p+2q-pq)}, \\
  \label{eq:N1_rpt}
  \frac{N_1}{N_3}&=\frac{n_1(p,q)}{r}
  =\frac{2pq}{r(2p+2q-pq)}, \\ 
  \label{eq:N2_rpt}
  \frac{N_2}{N_3}&=\frac{n_2(p,q)}{2}
  =\frac{2q}{2p+2q-pq},
\end{align}
and have a consistency with Schl\"afli's formula
\begin{align}
  \label{eq:schf}
  N_0-N_1+N_2-N_3=0.
\end{align}
Furthermore, it is known that 
$N_3$ is given by Coxeter's 
formula \cite{Coxeter, Hitotsumatsu}
\begin{align}
  \label{eq:N3_rpt}
  N_3
  =& \frac{ 32h_{pqr}}{pn_2(p,q)
    \left[12-p-2q-r+4\left(\dfrac{1}{p}+\dfrac{1}{r}\right)\right]},
\end{align}
where $h_{pqr}$ is a positive integer known as the Petrie number. It is related 
to the largest root of the quartic equation
\begin{align}
  \label{eq:petrie}
  x^4-\left(\cos^2\frac{\pi}{p}+\cos^2\frac{\pi}{q}
    +\cos^2\frac{\pi}{r}\right)x^2
  +\cos^2\frac{\pi}{p}\cos^2\frac{\pi}{r}=0
\end{align}
by $x=\cos\dfrac{\pi}{h_{pqr}}$. Equations (\ref{eq:N0_rpt})--(\ref{eq:N2_rpt}) and (\ref{eq:N3_rpt}) 
determine $N_{0,1,2,3}$. As we shall see, the ratios 
in Eqs (\ref{eq:N0_rpt})--(\ref{eq:N2_rpt}) are sufficient in writing the Regge 
equations. In Table \ref{tab:rpt} we summarize the properties of regular 
4-ploytopes for the reader's reference. 

\begin{figure}[t]
  \centering
  \includegraphics[scale=1]{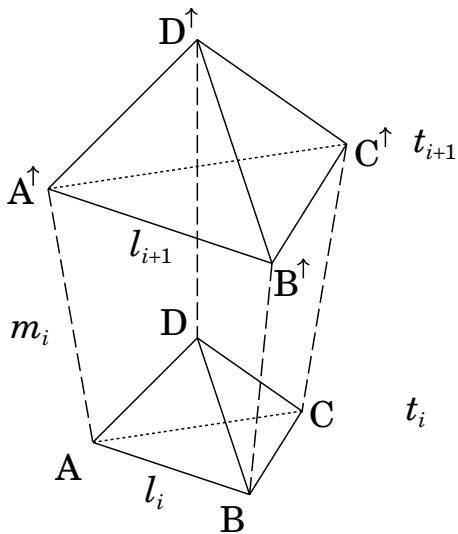}
  \caption{The $i$th frustum as the fundamental building block 
    of a 4-polytopal universe for $p=q=3$.
    A cell of regular tetrahedron ABCD with edge 
    length $l_i$ at time $t_i$ evolves into a cell 
    A$^\uparrow$B$^\uparrow$C$^\uparrow$D$^\uparrow$ with edge length $l_{i+1}$ 
    at $t_{i+1}$.}
  \label{fig:fbb}
\end{figure}

As depicted in Fig. \ref{fig:fbb}, the fundamental building blocks of 
space-time in the Regge calculus are world-tubes of four-dimensional 
frustums with the regular polyhedra $\{p,q\}$ as the upper and lower 
cells, and with $n_2 \left( p , q \right)$ lateral 
cells which are three-dimensional frustums with $p$-sided regular polygons 
as the upper and lower faces. In Fig. \ref{fig:fbb} the lateral cells are the three-dimensional frustums ABC-A$^\uparrow$B$^\uparrow$C$^\uparrow$, 
ABD-A$^\uparrow$B$^\uparrow$D$^\uparrow$, ACD-A$^\uparrow$C$^\uparrow$D$^\uparrow$, 
and BCD-B$^\uparrow$C$^\uparrow$D$^\uparrow$.

Following the regular polyhedron models \cite{TF:2016aa}, we assume that the lower  
and upper cells of a block separately lie in a time-slice and every strut between 
them has equal length. The whole space-time is then obtained by 
gluing such frustums cell-by-cell without a break. There are two types 
of fundamental variables: the lengths of the edges, $l_i$, and those of the 
struts, $m_i$. Since the hinges are two-dimensional faces in four dimensions, there are 
only two types of hinges. One is a face of a regular polyhedron in a time-slice, like 
$\triangle\mathrm{ABC}$ in Fig. \ref{fig:fbb}. We call 
it simply a ``polygon" and denote by $A^{({\rm p})}_i$ the area of the polygon on 
the $i$th Cauchy surface at time $t_i$. The other type of hinge is an isosceles 
trapezoidal face of lateral cells between the consecutive Cauchy surfaces, such as
$\square\mathrm{A}\mathrm{B}\mathrm{B}^\uparrow\mathrm{A}^\uparrow$ in Fig. \ref{fig:fbb}. 
We call them ``trapezoid'' and denote by $A^{({\rm t})}_i$ the area of the trapezoid 
between the Cauchy surfaces at $t_i$ and $t_{i+1}$. 

With these in mind, the Regge action (\ref{eq:ract}) can be written as
\begin{align}
  \label{eq:regact}
  S_\mathrm{Regge}=\frac{1}{8\pi}\sum_i \left(N_1A^{({\rm t})}_i\varepsilon_i^{\rm (t)}
  +N_2 A^{({\rm p})}_i\varepsilon_i^{\rm (p)}-N_3\Lambda V_i \right),
\end{align}
where $\varepsilon^{\rm ({\rm t})}_i$ and $\varepsilon_i^{\rm ({\rm p})}$ stand for the 
deficit angles around the trapezoid and polygon, respectively, and $V_i$ is the 
world-volume of the $i$th frustum. The sum on the right-hand side is taken over the 
time-slices. As we show in the next section, the deficit angles, areas, and volume 
are given in terms of the lengths of the edges and struts.




\section{Regge equations}

\label{sec:req}
\setcounter{equation}{0}

The Regge equations can be obtained by varying the action (\ref{eq:regact}) 
with respect to the fundamental variables $m_i$ and $l_i$. Note that two 
adjacent trapezoids $A^{({\rm t})}_i$ and $A^{({\rm t})}_{i-1}$ have the 
edge $l_i$ in common, as do $V_i$ and $V_{i-1}$. Then, Eq. (\ref{eq:regeq}) 
can be written as 
\begin{align}
  \frac{ \partial A^{\left( {\rm t} \right)}_i }{ \partial m_i } \varepsilon_i^{\left( {\rm t} \right)} 
  &= \frac{ r \left( 2p + 2q - pq \right) }{ 2 p q } 
  \Lambda \frac{ \partial V_i }{ \partial m_i } , 
  \label{eq:euc_hc_ptu} \\ 
  \frac{ \partial A^{ \left( {\rm t} \right) }_i }{ \partial l_i } \varepsilon_i^{ \left( {\rm t} \right) } 
  + \frac{ \partial A^{ \left( {\rm t} \right) }_{i-1} }{ \partial l_i } \varepsilon_{i-1}^{ \left( {\rm t} \right) } 
  &= - \frac{ r }{ p } \frac{ \partial A^{ \left( {\rm p} \right) }_i }{ \partial l_i } 
  \varepsilon^{ \left( {\rm p} \right) }_i + \frac{ r \left( 2p + 2q - pq \right) }{ 2 p q } \Lambda 
  \left( \frac{ \partial V_i }{ \partial l_i } + \frac{ \partial V_{i-1} }{ \partial l_i } \right).
\label{eq:euc_ev_ptu} 
\end{align}
In the context of the ADM formalism, 
Eq. (\ref{eq:euc_hc_ptu}) corresponds to the Hamiltonian constraint and Eq. (\ref{eq:euc_ev_ptu})
to the evolution equation. 

The deficit angles, areas of the hinges, and volume of the frustum can be expressed 
in terms of $l$ and $m$. For the sake of lucidness in defining lengths and 
angles, we temporally assume the metric in each building block to be flat 
Euclidean so that the geometric objects such as lengths and angles are obvious. 
The equations of motion in Lorentzian geometry can be achieved by Wick rotation. 

\begin{figure}[t]
  \centering
  \includegraphics[scale=0.8]{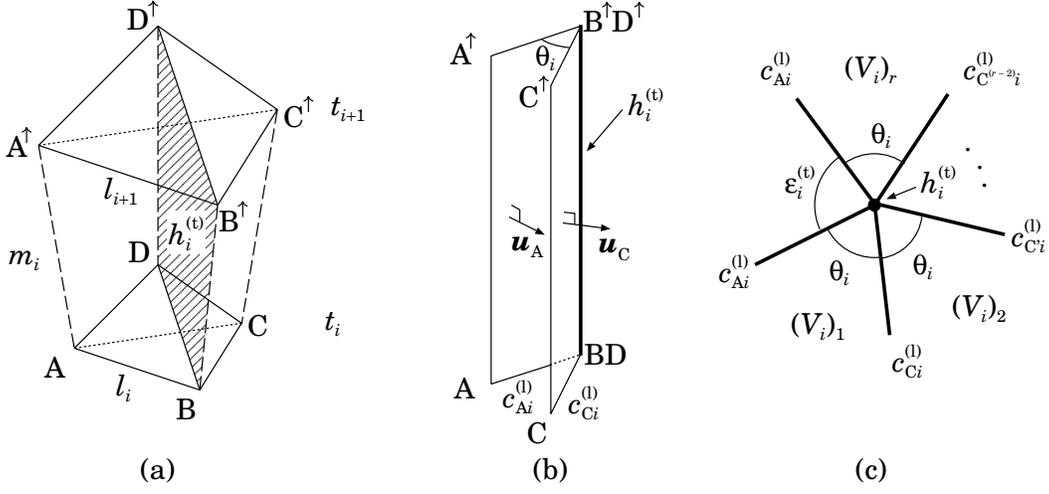}
  \caption{(a) two lateral cells $c^{(\mathrm{l})}_{\mathrm{A}i}$ and $c^{(\mathrm{l})}_{\mathrm{C}i}$ meeting 
    at the trapezoidal hinge $h^{(\mathrm{t})}_i$, (b) dihedral angle 
    between $c^{(\mathrm{l})}_{\mathrm{A}i}$ and $c^{(\mathrm{l})}_{\mathrm{C}i}$, and (c) deficit angle 
    around the hinge $h_i^{(\mathrm{t})}$ made by $r$ frustums $(V_i)_1$, $\cdots$, $(V_i)_r$ 
    having $h_i^{(\mathrm{t})}$ as a lateral 
    face in common. Though Figure (a) assumes an 
    evolution of a regular tetrahedron, any polyhedron can be used. }
  \label{fig:dha_trap}
\end{figure}

We first focus our attention on a trapezoidal hinge 
$h_i^{(\mathrm{t})}=\mathrm{BB}^\uparrow\mathrm{D}^\uparrow\mathrm{D}$ of the $i$th frustum, the shaded 
area of Fig. \ref{fig:dha_trap}(a). One sees that the two lateral cells 
$c^{(\mathrm{l})}_{\mathrm{A}i}=\mathrm{ABD}$-$\mathrm{A^\uparrow B^\uparrow D^\uparrow}$ and 
$c^{(\mathrm{l})}_{\mathrm{C}i}=\mathrm{BCD}$-$\mathrm{B^\uparrow C^\uparrow D^\uparrow}$ have the 
hinge $h_i^{(\mathrm{t})}$ in common as a face. We can find a unit normal 
vector $\boldsymbol{u}_{\mathrm{A}}$ to the cell $c^{(\mathrm{l})}_{\mathrm{A}i}$. It is orthogonal 
to vectors $\overrightarrow{\mathrm{BA}}$, $\overrightarrow{\mathrm{BD}}$, and 
$\overrightarrow{\mathrm{BB}}^\uparrow$. Similarly, we denote by $\boldsymbol{u}_{\mathrm{C}}$ a 
unit normal to the cell $c^{(\mathrm{l})}_{\mathrm{C}i}$. 
Then the dihedral angle $\theta_i$ between the two lateral cells is defined by 
\begin{align}
  \label{eq:thetai}
  \theta_i=\arccos\boldsymbol{u}_\mathrm{A}\cdot\boldsymbol{u}_\mathrm{C}. 
\end{align}
(See Fig. \ref{fig:dha_trap}b.) This is explicitly written as 
\begin{align}
  \theta_i=\arccos\frac{4\left(\sin^2\frac{\pi}{p}
      -2\cos^2\frac{\pi}{q}\right)m_i^2+\delta l_i^2
    \cos\frac{2\pi}{q}}{4m_i^2\sin^2\frac{\pi}{p}-\delta l_i^2},
  \label{eq:theta_rptu}
\end{align}
where $\delta l_i=l_{i+1}-l_i$. The deficit angle $\varepsilon_i^{({\rm t})}$ around 
the hinge $h^{(\mathrm{t})}_i$  can be found by noting the fact that there are $r$ 
frustums that have the trapezoid in common as illustrated 
in Fig. \ref{fig:dha_trap}(c). 
We thus obtain
\begin{align}
  \label{eq:datri} 
  \varepsilon_i^{({\rm t})}=2\pi-r\theta_i.
\end{align}

\begin{figure}[t]
  \centering
  \includegraphics[scale=0.7]{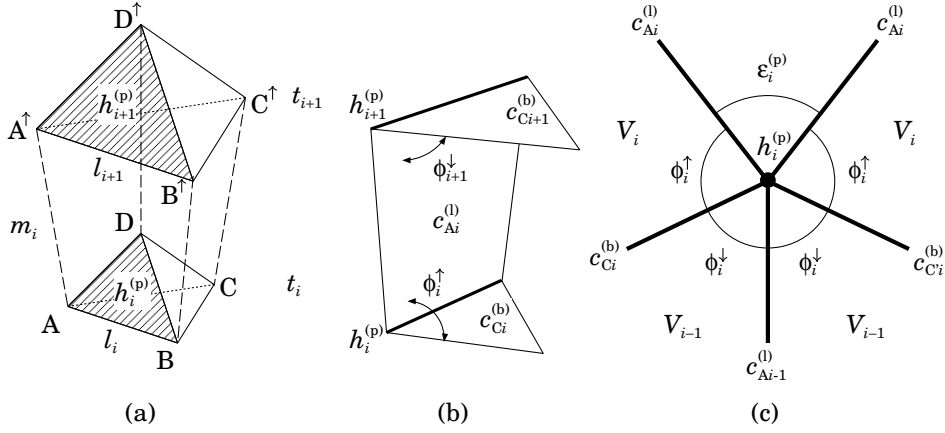}
  \caption{(a) two polygonal hinges $h_i^{(\mathrm{p})}$ and $h_{i+1}^{(\mathrm{p})}$ 
    in the $i$th frustum, (b) dihedral angles $\phi^\uparrow_i$ and $\phi^\downarrow_{i+1}$,
  and (c) deficit angle $\varepsilon_i^{({\rm p})}$. } 
  \label{fig:dha_plyg}
\end{figure}

We next pick up a pair of polygonal hinges $h^{(\mathrm{p})}_i=\mathrm{ABD}$ 
and $h^{(\mathrm{p})}_{i+1}=\mathrm{A^\uparrow B^\uparrow D^\uparrow}$ in the $i$th frustum 
as depicted in Fig. \ref{fig:dha_plyg}(a). They are the upper and lower faces of 
the lateral cell $c^{(\mathrm{l})}_{\mathrm{A}i}$ defined above. The lateral cell
$c^{(\mathrm{l})}_{\mathrm{A}i}$ and the base cell $c^{(\mathrm{b})}_{\mathrm{C}i}=\mathrm{ABCD}$ meet at 
the hinge $h^{(\mathrm{p})}_i$. We denote the dihedral angle between them by $\phi_i^\uparrow$. 
Similarly, we write the dihedral angle between $c^{(\mathrm{l})}_{\mathrm{A}i}$ and 
$c^{(\mathrm{b})}_{\mathrm{C}i+1}=\mathrm{A^\uparrow B^\uparrow C^\uparrow D^\uparrow}$
by $\phi^\downarrow_{i+1}$. Since $c^{(\mathrm{b})}_{\mathrm{C}i}$ and 
$c^{(\mathrm{b})}_{\mathrm{C}i+1}$ are parallel to each other, the dihedral angles 
satisfy
\begin{align}
  \phi^\uparrow_i+\phi^\downarrow_{i+1}=\pi. 
\end{align}
(See Fig. \ref{fig:dha_plyg}b.) The dihedral angle $\phi^\downarrow_i$ can 
be obtained by the way just explained for $\theta_i$ as  
\begin{align}
  \phi_i^\downarrow=\arccos\frac{\delta l_{i-1}
    \cos\frac{\pi}{p}\cos\frac{\pi}{q}}{\sqrt{\left(\sin^2\frac{\pi}{p}
        -\cos^2\frac{\pi}{q}\right)\left(4m_{i-1}^2\sin^2\frac{\pi}{p}-\delta l_{i-1}^2\right)}}.
\label{eq:phi_down_rptu}
\end{align}

To find the deficit angle around the hinge $h^{(\mathrm{p})}_i$, we must take account of 
four frustums that have $h_i^{(\mathrm{p})}$ in common: two adjacent $V_i$ in the future side and 
two adjacent $V_{i-1}$ in the past side, as schematically illustrated in 
Fig. \ref{fig:dha_plyg}(c). Then, the deficit angle $\varepsilon_i^{(\mathrm{p})}$ can 
be written as
\begin{align}
  \label{eq:daplyg}
  \varepsilon_i^{({\rm p})}=2 \pi-2 \left(\phi_i^{\uparrow}+\phi_i^{\downarrow} \right)
  =2\delta\phi_i^{\downarrow},
\end{align}
where we have introduced $\delta\phi_i^{\downarrow}
=\phi_{i+1}^{\downarrow}-\phi_i^{\downarrow}$. 

The areas of the hinges and the volume of the frustum can also be expressed in 
terms of $l$ and $m$. The areas of a trapezoid $A^{({\rm t})}_i$ and 
a polygon $A^{({\rm p})}_i$ can be written as
\begin{align}
A^{({\rm t})}_i&=\frac{1}{2}\left(l_{i+1}+l_i\right)\sqrt{m_i^2-\frac{1}{4}\delta l_i^2}, \\
A^{({\rm p})}_i&=\frac{p}{4}l_i^2\cot\frac{\pi}{p}.
\end{align}
The volume of the $i$th frustum is given by 
\begin{align}
  V_i=\frac{pq\cot^2\frac{\pi}{p}\cos\frac{\pi}{q}\left(l_{i+1}+l_i\right)
    \left(l_{i+1}^2+l_i^2\right)}{24\left(2p+2q-pq\right)
    \sqrt{\sin^2\frac{\pi}{p}-\cos^2\frac{\pi}{q}}}
  \sqrt{m_i^2-\frac{\sin^2\frac{\pi}{q}}{4\left(\sin^2\frac{\pi}{p}
        -\cos^2\frac{\pi}{q}\right)}\delta l_i^2}.
  \label{eq:fst_rptu} 
\end{align}
 Inserting these expressions into the Regge equations (\ref{eq:euc_hc_ptu}) and (\ref{eq:euc_ev_ptu}), we obtain a set of recurrence relations:
\begin{align}
\label{eq:euc_rr_hc_rptu} 
& \frac{ \varepsilon_i^{ \left( {\rm t} \right) } }{ \sqrt{ 4 m_i^2 - \delta l_i^2 } } = \frac{ r \Lambda}{ 24 }
  \frac{ \left( l_{i+1}^2 + l_i^2 \right) \cot^2 \frac{ \pi }{ p } \cos \frac{ \pi }{ q } }{%
    \sqrt{ 4 m_i^2 \left( \sin^2 \frac{ \pi }{ p } - \cos^2 \frac{ \pi }{ q } \right) - \sin^2 \frac{ \pi }{ q } \delta l^2_i } } , \\
  \label{eq:euc_rr_ev_rptu} 
\nonumber & \sqrt{ 4 m_i^2 - \delta l_i^2 } \varepsilon_i^{ \left( {\rm t} \right) } + \sqrt{ 4 m_{i-1}^2 - \delta l_{i-1}^2 } \varepsilon_{i-1}^{ \left( {\rm t} \right) } \\
\nonumber & \quad + \frac{ l_{i+1}^2 - l_i^2 }{ \sqrt{ 4 m_i^2 - \delta l_i^2 } } \varepsilon_i^{ \left( {\rm t} \right) } - \frac{ l_i^2 - l_{i-1}^2 }{ \sqrt{ 4 m_{i-1}^2 - \delta l_{i-1}^2 } } \varepsilon_{i-1}^{ \left( {\rm t} \right) } + 2 r \cot \frac{ \pi }{ p } l_i \varepsilon_i^{ \left( {\rm p} \right) } \\
\nonumber & = \frac{ r \Lambda \cot^2 \frac{ \pi }{ p } \cos \frac{ \pi }{ q } }{ 24 \left( \sin^2 \frac{ \pi }{ p } - \cos^2 \frac{ \pi }{ q } \right) } \Biggl[ \left( l_{i+1}^2 + 2 l_{i+1} l_i + 3 l_i^2 \right) \sqrt{ 4 m_i^2 \left( \sin^2 \frac{ \pi }{ p } - \cos^2 \frac{ \pi }{ q } \right) - \sin^2 \frac{ \pi }{ q } \delta l_i^2 } \\
\nonumber & \quad + \left( 3 l_i^2 + 2 l_i l_{i-1} + l_{i-1}^2 \right) \sqrt{ 4 m_{i-1}^2 \left( \sin^2 \frac{ \pi }{ p } - \cos^2 \frac{ \pi }{ q } \right) - \sin^2 \frac{ \pi }{ q } \delta l_{i-1}^2 } \\
& \quad + \frac{ \left( l_{i+1}^4 - l_i^4 \right) \sin^2 \frac{ \pi }{ q } }{ \sqrt{ 4 m_i^2 \left( \sin^2 \frac{ \pi }{ p } - \cos^2 \frac{ \pi }{ q } \right) - \sin^2 \frac{ \pi }{ q } \delta l_i^2 } } %
- \frac{ \left( l_i^4 - l_{i-1}^4 \right) \sin^2 \frac{ \pi }{ q } }{ \sqrt{ 4 m_{i-1}^2 \left( \sin^2 \frac{ \pi }{ p } - \cos^2 \frac{ \pi }{ q } \right) - \sin^2 \frac{ \pi }{ q } \delta l_{i-1}^2 } } \Biggr] .
\end{align}
These are non-linear recurrence relations for the edge and strut 
lengths $l_i$ and $m_i$. Evolution of the polytopal universe can be 
investigated by taking the continuum time limit as in Refs. 
\cite{CW:1973aa,LW:2015aa,LW:2015ab,LW:2015ac}.




\section{Continuum time limit}

\label{sec:ctl}
\setcounter{equation}{0}

We are interested in the evolution of a model universe with a regular polytope 
as the Cauchy surfaces. In continuum theory, Cauchy surfaces are defined as 
slices of space-time by constant times. In the FLRW universe, the time axis is 
taken to be orthogonal to the Cauchy surfaces. This seems to correspond to 
choosing the time axis to be orthogonal to the Cauchy cells. One can identify 
the distance between the centers of circumspheres of the two Cauchy cells in 
Fig. \ref{fig:fbb} with the Euclidean time interval $\delta t_i=t_{i+1}-t_i$. 
This works for Cauchy cells of regular polyhedrons \cite{LW:2015aa,LW:2015ab}. 
Later in this paper we consider Cauchy surfaces that are not necessarily 
regular polytopes. For general polytopal substitutions for 3-spheres as 
Cauchy surfaces, however, distances between two temporally consecutive Cauchy 
cells vary cell by cell. One cannot identify the temporal distances with 
a common time interval $\delta t_i$, as noted in Ref. \cite{TF:2016aa} for 
polyhedral universe.

We avoid the subtleties in identifying the time coordinate by supposing a 
fictitious point material in a state of rest spatially at each vertex of the polytopal Cauchy surface. 
Taking $t_i$ as the proper time of a clock standing by the fictitious 
material particle, we can identify strut length $m_i$ with the time interval 
\begin{align}
  \label{eq:midt}
  m_i=\delta t_i. 
\end{align}
The time axis is not defined to be orthogonal to the polytopal Cauchy 
surfaces. The orthogonality of the temporal axis with the spatial ones 
is restored in the continuum limit. 
We further choose all the time intervals $\delta t_i$ to be equal 
and then take the continuum time limit $\delta t_i \to dt$. 
The edge lengths can be regarded as a smooth function of time 
$l_i \to l(t)$, and 
\begin{align}
  \label{eq:ld}
  \delta l_i=\frac{\delta l_i}{\delta t_i}\delta t_i \quad\to\quad
  \dot ldt, 
\end{align}
where $\dot l=dl/dt$. It is straightforward to take the continuum 
time limit for Eqs. (\ref{eq:euc_rr_hc_rptu}) and (\ref{eq:euc_rr_ev_rptu}). 
We find
\begin{align}
  \label{eq:cehc_ptu} 
  & \frac{ \varepsilon^{ \left( {\rm t} \right) } }{ \sqrt{ 4 - \dot{l}^2 } } 
  = \frac{ r \Lambda }{ 12 } \frac{ l^2 \cot^2 \frac{ \pi }{ p } 
    \cos \frac{ \pi }{ q } }{ \sqrt{ 4 \left( \sin^2 \frac{ \pi }{ p } 
        - \cos^2 \frac{ \pi }{ q } \right) 
      - \dot{l}^2 \sin^2 \frac{ \pi }{ q } } }, \\
  \label{eq:ceev_ptu} 
  \nonumber & \sqrt{ 4 - \dot{l}^2 } \varepsilon^{ \left( {\rm t} \right) } 
  + \frac{ d }{ dt } \left( \frac{ l \dot{ l } }{ \sqrt{ 4 - \dot{l}^2 } } 
    \varepsilon^{ \left( {\rm t} \right) } \right) 
  + 2 r l \cot \frac{ \pi }{ p } \dot{\phi}^\downarrow \\
  \nonumber & = \frac{ r \Lambda }{ 12 } 
  \frac{ \cot^2 \frac{ \pi }{ p } \cos \frac{ \pi }{ q } }{%
    \sin^2 \frac{ \pi }{ p } - \cos^2 \frac{ \pi }{ q } } 
  \Biggl[ 3 l^2 \sqrt{ 4 \left( \sin^2 \frac{ \pi }{ p } 
      - \cos^2 \frac{ \pi }{ q } \right) 
    - \dot{l}^2 \sin^2 \frac{ \pi }{ q } } \\
  & \quad + \frac{ d }{ dt } \left( \frac{ l^3 \dot{ l } 
      \sin^2 \frac{ \pi }{ q } }{ \sqrt{ 4 \left( \sin^2 \frac{ \pi }{ p } 
          - \cos^2 \frac{ \pi }{ q } \right) 
        - \dot{l}^2 \sin^2 \frac{ \pi }{ q } } } \right) \Biggr] ,
\end{align}
where $\varepsilon^{({\rm t})}$ and $\dot{\phi}^\downarrow$ are, respectively, the 
continuum time limits of Eqs. (\ref{eq:datri}) and (\ref{eq:phi_down_rptu})
\begin{align}
\label{eq:cdat} 
\varepsilon^{ ( {\rm t} ) } &= 2 \pi - r \arccos \frac{ 4 \left( \sin^2 \frac{ \pi }{ p } - 2 \cos^2 \frac{ \pi }{ q } \right) + \dot{l}^2 \cos \frac{ 2 \pi }{ q } }{ 4 \sin^2 \frac{ \pi }{ p } - \dot{l}^2 } , \\
\label{eq:cdap} 
\dot{\phi}^\downarrow &= \frac{ d }{ dt } \arccos \frac{ \dot{l} \cos \frac{ \pi }{ p } \cos \frac{ \pi }{ q } }{ \sqrt{ \left( \sin^2 \frac{ \pi }{ p } - \cos^2 \frac{ \pi }{ q } \right) \left( 4 \sin^2 \frac{ \pi }{ p } - \dot{l}^2 \right) } } .
\end{align}
Since we have fixed the strut lengths by Eq. (\ref{eq:midt}), they disappear from 
the Regge equations. Furthermore, substituting Eqs. (\ref{eq:cehc_ptu}) 
and (\ref{eq:cdap}) into the evolution equation (\ref{eq:ceev_ptu}), 
it can be simplified as
\begin{align}
  \ddot{l}=-\frac{\Lambda}{3}l\left(1
    -\frac{\dot{l}^2}{4\sin^2\frac{\pi}{p}}\right)
  \left[1-\frac{1}{4}\dot{l}^2
    +\frac{1}{2}\frac{l\ddot{l}\cos^2\frac{\pi}{p}}{4\left(\sin^2\frac{\pi}{p}
        - \cos^2\frac{\pi}{q}\right)-\dot{l}^2\sin^2\frac{\pi}{q}}\right].
  \label{eq:ceev_ptu_simp} 
\end{align}
One can easily verify that  this is consistent with the Hamiltonian 
constraint (\ref{eq:cehc_ptu}). In other words, the Hamiltonian constraint (\ref{eq:cehc_ptu}) 
can be obtained as the first integral of the evolution equation (\ref{eq:ceev_ptu_simp}) for the initial conditions  
\begin{align}
l \left( 0 \right) = l_0 = \sqrt{ \frac{ 12 }{ r \Lambda } \left( 2 \pi - r \theta_0 \right) \cot \frac{ \theta_0 }{ 2 } } \tan \frac{ \pi }{ p } , \quad \dot{l} \left( 0 \right) = 0 ,
\label{eq:icohcfrptu} 
\end{align}
where $\theta_0 = 2 \arcsin \left[ \cos \left( \pi / q \right) / \sin \left( \pi / p \right) \right]$ stands for a dihedral angle of the regular polyhedron $ \{ p , q \} $. The cosmological constant must be positive for regular 4-polytopes 
as we see from Eq. (\ref{eq:icohcfrptu}). This implies that the space-time 
is de Sitter-like. The 4-polytopal universe cannot expand from or contract to a point but has minimum edge length $l_0$, as does the continuum solution, as we shall see below.

So far we have worked with piecewise linear space-time with Euclidean 
signature. To argue the evolution of 
space-time we move to the Minkowskian signature by Wick rotation. 
This can be done simply by letting  $\dot l^2, \ddot l 
\to -\dot l^2, -\ddot l$ in Eqs. (\ref{eq:cehc_ptu}) 
and (\ref{eq:ceev_ptu_simp}). We thus obtain 
\begin{align}
  \label{eq:chc_ptu}
  \nonumber & 2\pi-r\arccos\frac{4\left(\sin^2\frac{\pi}{p}
      -2\cos^2\frac{\pi}{q}\right)
    -\dot{l}^2\cos\frac{2\pi}{q}}{4\sin^2\frac{\pi}{p}+\dot{l}^2}\\
  & ~ =\frac{r\Lambda}{12} l^2
  \sqrt{\frac{4+\dot{l}^2}{4\left(\sin^2\frac{\pi}{p}
        -\cos^2\frac{\pi}{q}\right)
      +\dot{l}^2\sin^2\frac{\pi}{q}}}\cot^2\frac{\pi}{p}\cos\frac{\pi}{q} , \\
  \label{eq:cev_ptu}
  &\ddot{l} =\frac{\Lambda}{3} l\left(1
    +\frac{\dot{l}^2}{4\sin^2\frac{\pi}{p}}\right)
  \left[ 1+\frac{1}{4}\dot{l}^2
    -\frac{1}{2}\frac{l\ddot{l}\cos^2\frac{\pi}{p}}{4
      \left(\sin^2\frac{\pi}{p}-\cos^2\frac{\pi}{q}\right)
      +\dot{l}^2\sin^2\frac{\pi}{q}}\right] .
\end{align}
From the evolution equation (\ref{eq:cev_ptu}) we see that the 
acceleration $ \ddot{l} $ is always positive. Hence the polytopal 
universe, as the continuum solution, exhibits accelerated expansion 
or decelerated contraction with the minimum edge length 
(\ref{eq:icohcfrptu}). The universe, however, reaches a 
maximum size in a finite period of time as we shall see in 
the next section. 

As a consistency check, let us consider the case of a vanishing 
cosmological constant before turning to a detailed exposition of the 
behavior of the polytopal universe described by the evolution 
equation (\ref{eq:cev_ptu}). 
In the absence of a cosmological constant, the Hamiltonian constraint 
(\ref{eq:chc_ptu}) becomes
\begin{align}
  \dot{l}^2 = \frac{8\left(\sin^2\frac{\pi}{p}\sin^2\frac{\pi}{r}
      -\cos^2\frac{\pi}{q}\right)}{\cos\frac{2\pi}{q}+\cos\frac{2\pi}{r}} 
  = \mbox{const.} \geq 0 .
\label{eq:hcwc_ptu} 
\end{align}
There is no convex regular 4-polytope that has a Schl\"afli symbol 
satisfying this inequality. In the case of $\dot{l}^2 = 0$, it 
admits $ \left\{ p , q , r \right\} = \left\{ 4 , 3, 4 \right\} $ which gives a flat 
Cauchy surface corresponding to the Minkowski metric. Moreover, in 
the case of $\dot{l}^2 > 0$, a Schl\"afli symbol satisfying this 
inequality stands for a regular lattice of open Cauchy surface of 
constant negative curvature. These results are consistent with 
solutions of the Friedmann equations (see Table \ref{table:4d_flrw}).




\section{Numerical solution}

\label{sec:nsol}
\setcounter{equation}{0}

In this section, we solve the Hamiltonian constraint (\ref{eq:chc_ptu}) 
numerically and examine the behaviors of the regular 4-polytopal universes.
It is convenient to use the continuum time limit of the dihedral 
angle (\ref{eq:theta_rptu}). 
Let us denote it by $\theta$:
\begin{align}
\theta = \arccos \frac{ 4 \left( \sin^2 \frac{ \pi }{ p } - 2 \cos^2 \frac{ \pi }{ q } \right) - \dot{l}^2 \cos \frac{ 2 \pi }{ q } }{ 4 \sin^2 \frac{ \pi }{ p } + \dot{l}^2 }.
\label{eq:theta_cnt_ptu} 
\end{align}
Then $l$ and $\dot l$ can be expressed as 
\begin{align}
\label{eq:ld^2_ptu}
\dot{l}^2 &= \frac{ 4 \left( \cos^2 \frac{ \pi }{ q } - \sin^2 \frac{ \pi }{ p } \sin^2 \frac{ \theta }{ 2 } \right) }{ \sin^2 \frac{ \theta }{ 2 } - \cos^2 \frac{ \pi }{ q } } , \\
\label{eq:l^2_ptu}
l^2 &= \frac{ 12 }{ r \Lambda } \left( 2 \pi -r \theta \right) \tan^2 \frac{ \pi }{ p } \cot \frac{ \theta }{ 2 }.
\end{align}
The first of these can be obtained directly from Eq. (\ref{eq:theta_cnt_ptu}). 
The second can be derived from the Hamiltonian constraint (\ref{eq:chc_ptu}) 
by replacing $\dot l^2$ with Eq. (\ref{eq:theta_cnt_ptu}). Since 
$\dot l^2\geq0$, the dihedral angle varies in the range 
$\theta_q \leq \theta \leq \theta_0$, where $\theta_q=(q-2)\pi/q$. 
The velocity $\dot l$ diverges for $\theta=\theta_q$, while 
the edge length $l$ approaches a finite value $l=l_{p,q,r}$, where 
\begin{align}
  \label{eq:lpqr}
  l_{p,q,r}=\sqrt{\frac{12\pi}{\Lambda}\left(\frac{2}{q}+\frac{2}{r}-1\right)
    \tan\frac{\pi}{q}}\tan\frac{\pi}{p}.
\end{align}
This is contrasted with the polyhedral universe \cite{TF:2016aa},
where both $l$ and $\dot l$ diverge at a finite time. 

\begin{figure}[t]
  \centering
  \includegraphics[scale=0.8]{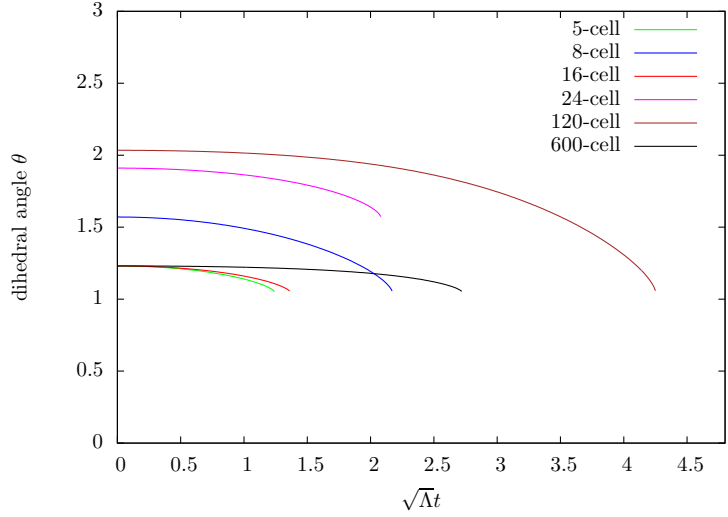}
  \caption{Plots of the dihedral angles of the regular 4-polytope models.
    Each plot ends at $t=\tau_{p,q,r}/2$. }
  \label{fig:rpt_dha}
\end{figure}

To see these in more detail we eliminate the edge length from 
Eqs. (\ref{eq:ld^2_ptu}) and (\ref{eq:l^2_ptu}) to obtain 
\begin{align}
\dot{ \theta } = \mp \frac{ 2 \sin \frac{ \theta }{ 2 } \csc \frac{ \theta_0 }{2} }{ 2 \pi - r \left( \theta - \sin \theta \right) }\sqrt{ \frac{ 2 r \Lambda }{ 3 } \left( 2 \pi - r \theta \right) \sin \theta \frac{ \left( \sin^2 \frac{\theta_0}{ 2 } - \sin^2 \frac{\theta_q}{ 2 } \right) \left( \sin^2 \frac{ \theta_0 }{ 2 } - \sin^2 \frac{ \theta }{ 2 } \right) }{ \sin^2 \frac{ \theta }{ 2 } - \sin^2 \frac{ \theta_q }{ 2 } } } .
\label{eq:tde_ptu} 
\end{align}
where the upper (lower) sign corresponds to an expanding (contracting) universe.  
Integrating Eq. (\ref{eq:tde_ptu}) numerically for the initial condition
\begin{align}
\theta \left( 0 \right) = \theta_0,
\label{eq:icotfpcu} 
\end{align}
we obtain numerical solutions for the dihedral angle. In Fig. \ref{fig:rpt_dha} 
we give the plots of dihedral angles for the six types of regular 4-polytopes. 
They are monotone decreasing functions of time for $0\leq t\leq \tau_{p,q,r}/2$ 
and approach $\theta_q$ as 
$t\to\tau_{p,q,r}/2$, where $\tau_{p,q,r}$ is defined by 
\begin{align}
  \tau_{p,q,r}=2\int^{\theta_0}_{\theta_q}d\theta
  \frac{2\pi-r\left(\theta-\sin\theta\right)}{ 2\sin\frac{\theta}{2} \csc \frac{ \theta_0 }{2} }
  \sqrt{\frac{ 3 \left(\sin^2\frac{\theta}{2}
        -\sin^2\frac{\theta_q}{2}\right)}{2 r \Lambda \left(2\pi-r\theta\right)
      \left( \sin^2 \frac{\theta_0}{ 2 } - \sin^2 \frac{\theta_q}{ 2 } \right) \left(\sin^2\frac{\theta_0}{2}-\sin^2\frac{\theta}{2}\right)\sin\theta}}.
      \label{eq:tpqr}
\end{align}
As noted above, the edge length becomes the maximum value 
$l=l_{p,q,r}$ at $t=\tau_{p,q,r}/2$. 
After reaching $l_{p,q,r}$, one of the most reasonable and easiest way is 
that the edge length begins to decrease.
The differential equation for the dihedral angle (\ref{eq:tde_ptu}) changes its sign to positive at the terminal time $ \tau_{p,q,r} / 2 $,
then the universe begins 
to contract. It continues until the edge length reaches 
the initial minimum value $l=l_0$ at $t=\tau_{p,q,r}$. The universe 
begins to expand again, repeating expansion and contraction with 
a period $\tau_{p,q,r}$. 
On the condition (\ref{eq:icohcfrptu}), the velocity $\dot{l}$ diverges at the time $\tau_{p,q,r} / 2$.
The edge length $l$ varies in the range $l_0 \leq l \leq l_{p,q,r}$.
Therefore there is no smooth continuation to the scale factor for $t \geq \tau_{p,q,r}/2$.
If we employ another initial condition for the evolution equation, 
positions of spikes are shifted, but do not disappear.
Thus the spikes surely come in the time evolution of the scale factor.

\begin{figure}[t]
  \centering
  \includegraphics[scale=0.8]{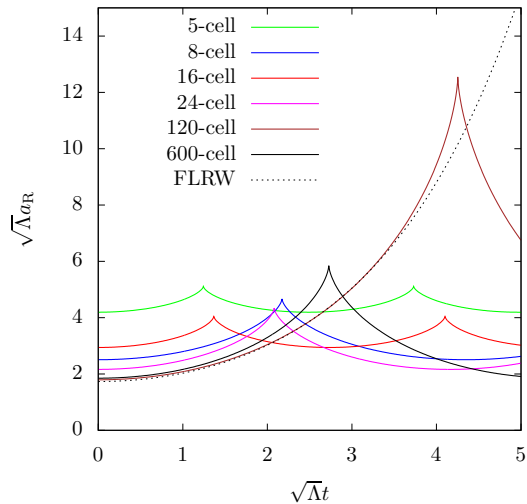}
  \caption{Plots of the scale factors of the regular 4-polytope models.}
  \label{fig:sfrpt}
\end{figure}

For comparison with the continuum theory, we introduce an analog of the 
scale factor $a(t)$. Here, we simply define the scale factor of the 
polytopal universe $a_\mathrm{R}$ as the radius of the circumsphere 
of the regular polytope
\begin{align}
  a_{\rm R}\left(t\right)=&\frac{l\left(t\right)}{2}
  \sqrt{\frac{\sin^2\frac{\pi}{q}-\cos^2\frac{\pi}{r}}{\sin^2\frac{\pi}{p}
      \sin^2\frac{\pi}{r}-\cos^2\frac{\pi}{q}}} \nonumber \\
  =&\sqrt{\frac{3\cot\frac{\theta}{2}\left(2\pi-r\theta\right)
      \left(\sin^2\frac{\pi}{r}-\sin^2\frac{\theta_q}{2}\right)}{%
      r\Lambda\left(\sin^2\frac{\theta_0}{2}-\sin^2\frac{\theta_q}{2}\right)
      \left(\sin^2\frac{\pi}{r}-\sin^2\frac{\theta_0}{2}\right)}}\sin\frac{\theta_0}{2}. 
  \label{eq:aRt} 
\end{align}
In Fig. \ref{fig:sfrpt} we give the plots of the scale factors of the regular 
4-polytopal universes as functions of time. The broken curve corresponds to the 
continuum solution. One can see that the regular 4-polytopal solutions 
approximate the continuum solution for $\sqrt{\Lambda} t < 1$. 
In particular, the more vertices the 
polytope contains, the better approximation is achieved.

That the numerical solutions reproduce the behavior of the 
continuum solution well for $\sqrt{\Lambda} t < 1$ can be understood 
by noting the fact that the scale factor (\ref{eq:aRt}) approximately 
satisfies the Friedmann equations (\ref{eq:Feq}) when both $\sqrt{\Lambda}l$ 
and $\dot l$ are small. In fact, the Hamiltonian constraint 
(\ref{eq:chc_ptu}) and the evolution equation (\ref{eq:cev_ptu}) can be 
approximated as
\begin{align}
  \label{eq:reldz_ptu} 
  \frac{ \Lambda }{ 3 } l^2 - \dot{l}^2= \frac{ \Lambda }{ 3 } l_0^2, 
  \qquad
  \ddot{l}= \frac{ \Lambda }{ 3 } l .
\end{align}
The deviations from the continuum solution, however, get large 
with time. In particular the polytopal universes 
repeat expansion and contraction. This is a unique 
characteristic of the polytope models. It cannot be seen in 
the FLRW universe or polyhedral universes. 

\section{Noninteger Schl\"afli symbol and pseudo-regular 4-polytopal universes}

\label{sec:prpt} 
\setcounter{equation}{0}

In the previous section, we investigated the behaviors of the regular 
4-polytopal universes. We have shown that the universes repeat expansion 
and contraction periodically. This property cannot be seen in 
an FLRW universe of the continuum general relativity. In the rest of this paper we are concerned 
with the issue of whether the model recovers the FLRW universe in the continuum 
limit, or not. A straightforward way to implement the continuum limit is 
to introduce an extension of the geodesic domes we have considered for polyhedron
models \cite{TF:2016aa}. 

\begin{figure}[t]
  \centering
  \includegraphics[scale=0.6]{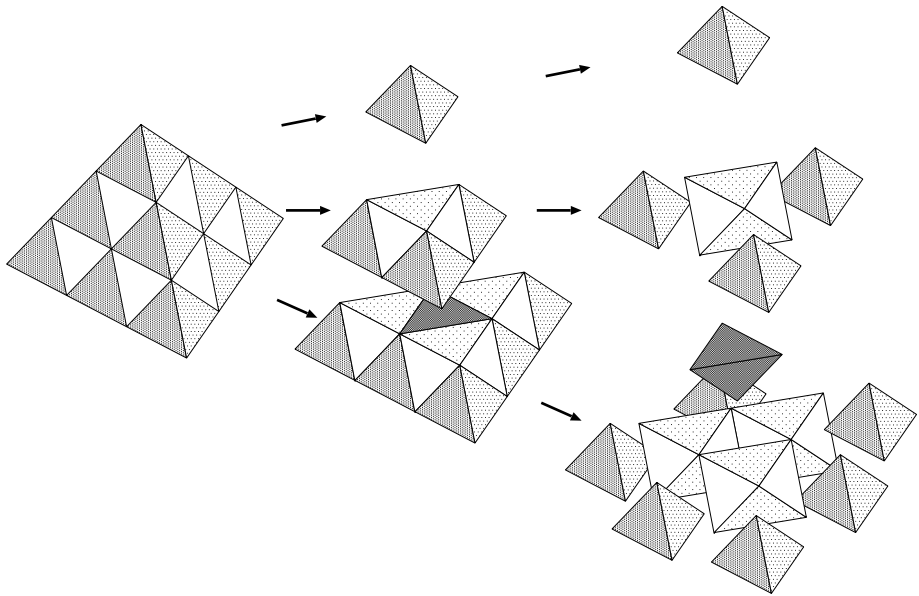}
  \caption{Subdivision of a regular tetrahedron in the case of $\nu = 3$.}
  \label{fig:dectetra}
\end{figure}

To better approximate a sphere beyond regular polyhedra we considered 
geodesic domes in Ref. \cite{TF:2016aa}. We now extend them to 
4-polytopes. To define a four-dimensional geodesic dome, geodesic 4-dome 
in brief, we divide each cell of a regular polytope into smaller polyhedra. 
By projecting the vertices of the subdivided cells on to the circumsphere of 
the original regular polytope we can define a geodesic 4-dome, a polytope 
having the points projected on to the circumsphere as the vertices. The 
method of subdivision is rather arbitrary and depends on the regular 
polyhedron to be subdivided. In Ref. \cite{Brewin:1987aa}, Brewin 
proposed a subdivision of a tetrahedron into tetrahedra that are 
not regular. 
Here we require that the subdivision of a 
regular polyhedron should yield regular polyhedra of equal edge 
length. In fact, a cube can be subdivided into $\nu^3$ smaller cubes 
of edge of a one-$\nu$th edge length, where $\nu$ is a positive 
integer called ``frequency''. Similarly, a regular tetrahedron and 
octahedron can be subdivided into smaller regular tetrahedra and 
octahedra as illustrated in Fig. \ref{fig:dectetra} for a regular 
tetrahedron. We can then construct geodesic 4-domes by applying 
these subdivisions to regular polytopes except for 120-cell. 
Since a dodecahedron has no subdivision into smaller regular 
polyhedra, we will not consider geodesic 4-domes for 120-cell.

Regge calculus for the geodesic 4-domes becomes cumbersome as the 
frequency $\nu$ increases as we have shown in Ref. \cite{TF:2016aa} 
in three dimensions. We can avoid this by regarding the geodesic 
4-domes as pseudo-regular 4-polytopes described by a non-integer 
Schl\"afli symbol. In what follows we consider as the polytopal 
universe pseudo-regular 4-polytopes corresponding to 600-cell-based 
geodesic 4-domes. We first define the Schl\"afli symbol characterizing 
pseudo-regular 4-polytopes. 

For our purpose we summarize the numbers of cells, faces, edges, and 
vertices of the geodesic 4-dome in Table \ref{table:4dim_gd_prop}.
At a frequency $\nu$ each cell of 
a 600-cell can be subdivided into $\nu(\nu^2+2)/3$ 
tetrahedra and $\nu(\nu^2-1)/6$ octahedra as depicted in 
Fig. \ref{fig:dectetra}. The geodesic 4-dome is then obtained by 
projecting the 600-cell tessellated by the $300(\nu^2+1)$ tiles on to 
the circumsphere. 
It has three types of triangular faces: one is a common face of two 
tetrahedra, another shared by two octahedra, and the other connecting 
a tetrahedron with an octahedron. Let us call them ``Tetra-Tetra 
connectors'', ``Octa-Octa connectors'', and ``Tetra-Octa connectors'', 
respectively. Furthermore, there are two types of edges. 
One is shared by five cells. They are coming from the edges of the 
original regular 600-cell. The other type of edges is shared by 
four cells. They correspond to newly generated edges by subdividing 
tetrahedral cells. Let us call the former type of edges ``five-way 
connectors'' and the latter ``four-way connectors'', respectively. 

\begin{table}[t]
\centering
\begin{tabular}{ccc}
\hline 
Frequency & & $ \nu $ \\ \hline
            & Tetrahedra & $ 200 \nu \left( \nu^2 + 2 \right) $ \\
$N_3$ & Octahedra & $ 100 \nu \left( \nu^2 - 1 \right) $ \\
            & Total & $ 300 \nu \left( \nu^2 + 1 \right) $ \\
            \ \\
            & Tetra-Tetra connectors & $ 600 \nu \left( \nu + 1 \right) $  \\
$N_2$ & Octa-Octa connectors & $ 600 \nu \left( \nu - 1 \right) $ \\
           & Tetra-Octa connectors & $ 400 \nu \left( 2 \nu^2 - 3 \nu + 1 \right) $ \\
           & Total & $ 400 \nu \left( 2 \nu^2 + 1 \right) $ \\
           \ \\
           & Five-way connectors & $ 720 \nu $ \\
$N_1$ & Four-way connectors & $ 600 \nu \left( \nu^2 - 1 \right) $ \\
           & Total & $ 120 \nu \left( 5 \nu^2 + 1 \right) $ \\
           \ \\
$N_0$ & & $ 20 \nu \left( 5 \nu^2 + 1 \right) $ \\ \hline
\end{tabular}
\caption{Numbers of cells, faces, edges, and vertices of a four-dimensional geodesic dome.}
\label{table:4dim_gd_prop}
\end{table}

A naive way to define the Schl\"afli symbol $\{p,q,r\}$ for the pseudo-regular 
polytope is taking the average number of edges of a face as $p$, of faces 
around a vertex in a cell as $q$, and of cells around an edge as $r$. This leads to 
\begin{align}
   \label{eq:ssprpt} 
   \left\{p,q,r\right\}=\left\{3,\frac{12\left(2\nu^2+1\right)}{7\nu^2+5}, 
  \frac{10\left(2\nu^2+1\right)}{5\nu^2+1}\right\}.
\end{align}
Pseudo-regular polytopes corresponding to the fractional Schl\"afli 
symbol (\ref{eq:ssprpt}), however, do not approach the three-dimensional sphere $S^3$ 
in the infinite frequency limit. This can be seen by noting the fact 
that a deficit angle $\varepsilon$ around an edge of a regular 
polytope is given by $\varepsilon=2\pi-2r\arcsin
\left[\cos \left( \pi/q \right)/\sin \left( \pi/p \right) \right]$. When applied 
to the pseudo-regular polytope, the deficit angle should satisfy 
$\varepsilon\rightarrow0$ for $\nu\rightarrow\infty$ or 
$\left\{p,q,r\right\}~\to~\left\{3,24/7,4\right\}$. It is obvious that this is not 
the case. Such a discrepancy happens since the subdivision of a cell involves 
two types of polyhedra: tetrahedra and octahedra. If we considered 
an 8-cell-based geodesic 4-dome, we could obtain a pseudo-regular polytope 
characterized by a fractional Schl\"afli symbol having a limit
$\{p,q,r\}~\to~\{4,3,4\}$. 

What we have shown is that the idea of averaging the Schl\"afli 
symbols does not work except for an 8-cell-based geodesic 4-dome. 
We must employ another method of averaging 
to define Schl\"afli symbols that not only have a smooth continuum 
limit but also preserve the geometrical characteristics of 
polytopes. 

To achieve this goal we introduce a set of angles 
$\vartheta_2$, $\vartheta_3$, and $\vartheta_4$, where $\vartheta_2$ 
is an interior angle of a face of a regular polytope $\{p,q,r\}$, 
$\vartheta_3$ a dihedral angle of two adjacent faces,
and $\vartheta_4$ a hyperdihedral angle between two neighboring 
cells. They can be written in terms of $p$, $q$, and $r$ as 
\begin{align}
  \vartheta_2&=\frac{p-2}{p}\pi=2\arcsin\left(\cos\frac{\pi}{p}\right), \\
  \vartheta_3&=2\arcsin
  \left(\frac{\cos\frac{\pi}{q}}{\sin\frac{\pi}{p}}\right),\\
  \vartheta_4&=2\arcsin
  \left(\frac{\sin\frac{\pi}{p}
      \cos\frac{\pi}{r}}{\sqrt{\sin^2\frac{\pi}{p}-\cos^2\frac{\pi}{q}}}\right).
\end{align}
These can be solved with respect to the Schl\"afli symbol as 
\begin{align}
  \label{eq:p_da}
  p(\vartheta_2)&=\frac{\pi}{\arccos\left(\sin\frac{\vartheta_2}{2}\right)},\\
  \label{eq:q_da}
  q(\vartheta_2,\vartheta_3)&
  =\frac{\pi}{\arccos\left(\cos\frac{\vartheta_2}{2}
      \sin\frac{\vartheta_3}{2}\right)},\\
  \label{eq:r_da}
  r(\vartheta_3,\vartheta_4)&=\frac{\pi}{\arccos
    \left(\cos\frac{\vartheta_3}{2}\sin\frac{\vartheta_4}{2}\right)}.
\end{align}

We are now able to extend the Schl\"afli symbol to an arbitrary 
pseudo-regular polytope by substituting in Eqs. (\ref{eq:p_da})--(\ref{eq:r_da})
the angles $\vartheta_{2,3,4}$
with averaged ones of tessellated parent regular polytopes. 
With the help of Table \ref{table:4dim_gd_prop}, it is straightforward 
to obtain the averaged angles of tessellated 600-cell as 
\begin{align}
  \label{eq:ada2_prpt} 
  \vartheta_2&=\frac{\pi}{3}, \\ 
  \label{eq:ada3_prpt} 
  \vartheta_3&=\frac{\left(\nu^2-1\right)\pi+3\arccos\frac{1}{3}}{2\nu^2+1}, \\ 
  \label{eq:ada4_prpt} 
  \vartheta_4&=\frac{\left(2\nu^2-3\nu+1\right)\pi
    +3\nu\arccos\left(-\frac{3\sqrt{5}+1}{8}\right)}{2\nu^2+1}.
\end{align}
The pseudo-regular 4-polytope with the Schl\"afli symbol 
(Eqs. \ref{eq:p_da}--\ref{eq:ada4_prpt}) has a smooth $S^3$ 
limit for $\nu\rightarrow\infty$ since 
$\left\{p,q,r\right\} \to \left\{3,\pi/\arccos \left( \sqrt{6}/4 \right) ,4\right\}$ and, hence, 
the deficit angle around an edge of a pseudo-regular 4-polytope 
satisfies $ \varepsilon=2\pi-2r\arcsin \left[ \cos \left(\pi/q \right)/\sin \left( \pi/p \right) \right]
\to 0$.

\begin{figure}[t]
  \centering
  \includegraphics[scale=0.8]{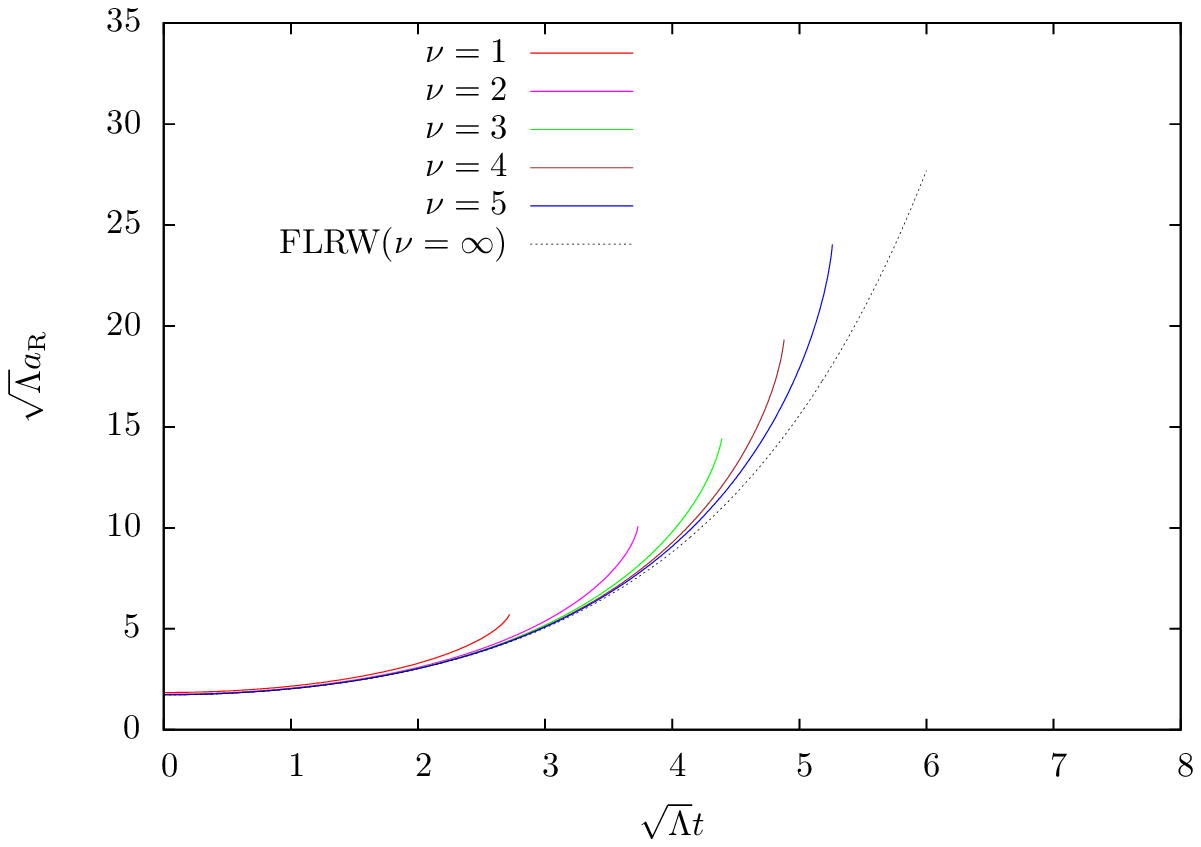}
  \caption{Plots of the scale factors of the pseudo-regular 4-polytope models for $\nu \leq 5$.}
  \label{fig:prptusf} 
\end{figure}

\begin{figure}[t]
  \centering
  \includegraphics[scale=0.8]{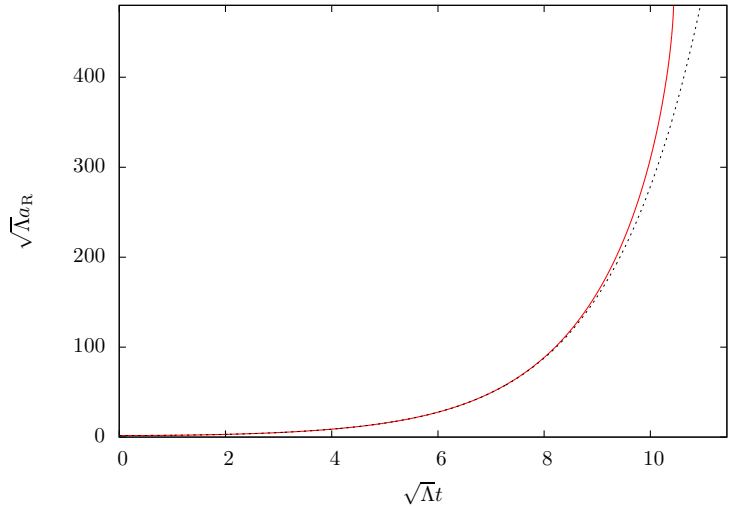}
  \caption{Plot of the scale factor of the pseudo-regular 4-polytope model for $\nu =100$. The broken curve is corresponding to the continuum FLRW universe.}
  \label{fig:prptusf_nu100} 
\end{figure}

Assuming that the expression for the scale factor (\ref{eq:aRt}) 
can be applied to the pseudo-regular 4-polytopal universe with the 
non-integer Schl\"afli symbol defined by 
Eqs. (\ref{eq:p_da})--(\ref{eq:ada4_prpt}), we can immediately 
obtain numerical solutions for an arbitrary frequency $\nu$. 
In Fig. \ref{fig:prptusf} we give plots of the scale factors 
for $\nu \leq 5$ and $0\leq t\leq \tau_{p,q,r}/2$. It can easily be 
seen that the pseudo-regular 4-polytope model 
approaches the FLRW universe as the frequency increases. 
The universe oscillates periodically for finite frequencies but 
the maximum scale factor and the period of oscillation grow with 
$\nu$. The scale factor for $\nu=100$ is shown in 
Fig. \ref{fig:prptusf_nu100}. That the Regge equations for the 
pseudo-regular 4-polytope reproduce the Friedmann equations (\ref{eq:Feq})
in the infinite frequency limit can be directly seen by substituting 
the scale factor (\ref{eq:aRt}) for the Hamiltonian constraint 
(\ref{eq:chc_ptu}) and the evolution equation (\ref{eq:cev_ptu}), 
and then taking the limit 
$\left\{p,q,r\right\}\to\left\{3,\pi/\arccos \left(\sqrt{6}/4\right),6\right\}$.




\section{Summary and discussions}

\label{sec:sum}
\setcounter{equation}{0}

We have investigated a four-dimensional closed FLRW universe with 
a positive cosmological constant using the CW formalism in Regge 
calculus. The main objective of this work is to extend the 
approach developed in Ref. \cite{TF:2016aa} for three-dimensional models 
to four dimensions. In particular, generalization of the method of 
pseudo-regular polyhedra and fractional Schl\"afli symbol is our 
prime concern. 

The three-dimensional hyperspherical Cauchy surfaces of 
the continuum FLRW universe are replaced with regular 4-polytopes 
instead of regular polyhedra in three-dimensional models. In four 
dimensions, the curvature reveals itself as deficit angles around 
hinges of two-dimensional faces. This makes the task in carrying out 
Regge calculus a bit cumbersome. We can handle the matter for 
all six types of regular 4-polytopes in a unified way in terms 
of the Schl\"afli symbol. We have seen that discretized 
universe changes the behavior according to the space-time 
dimensions. In the case of polyhedral universe 
the scale factor becomes infinite in a finite time. This is 
because the dihedral angle corresponding to Eq. (\ref{eq:theta_cnt_ptu}) 
approaches zero in a finite time as the universe expands to infinity. 
In four dimensions the polytopal universe stops expanding with a 
finite edge length in a finite time and then begins to contract to 
the scale where the universe starts expansion again. We thus arrive
at a picture of an oscillating universe. 

To go beyond the regular polytopes we have introduced pseudo-regular 
polytopes and a non-integer Schl\"afli symbol as a substitute of 
geodesic 4-domes. The fractional Schl\"afli symbol introduced in 
Ref. \cite{TF:2016aa} for the pseudo-regular polyhedra is defined by 
averaging the data such as the number of faces that have a vertex in 
common. When applied to pseudo-regular polytopes, the fractional 
Schl\"afli symbol does not recover a smooth three-dimensional 
sphere in the infinite frequency limit. 
To get rid of this 
difficulty we have employed averaged dihedral angles to define the 
Schl\"afli symbol by using Eqs. (\ref{eq:p_da})--(\ref{eq:r_da}). 
It is neither integral nor fractional. We have shown that the 
continuum FLRW universe can be reproduced in the infinite 
frequency limit of the oscillating polytopal universe. 

Our concern in this work is restricted to the vacuum solution 
of Einstein gravity with a positive cosmological constant. 
Furthermore, we have assumed compact hyperspherical Cauchy 
surfaces corresponding to positive curvature parameter. 
Inclusion of non-hyperspherical Cauchy surfaces as well as 
gravitating matter would certainly be interesting. 
We can also apply the approach of pseudo-regular polytopes to
a higher-dimensional FLRW universe. In $D(>4)$ dimensions 
there are only three types of regular polytopes: $D$-simplex, 
$D$-cube, and  $D$-orthoplex, as mentioned in Appendix 
\ref{sec:crrdp}. We expect that this makes the Regge calculus 
of polytopal universes in five or more dimensions easier. 

\vskip .5cm



\begin{center}
  {\bf Acknowledgments}
\end{center}
The authors would like to thank Y. Hyakutake, N. Motoyui, 
M. Sakaguchi, and S. Tomizawa for useful discussions.




\appendix

\section{Circumradius of a regular $D$-polytope}

\label{sec:crrdp}

For a regular 4-polytope $ \left\{ p , q , r \right\} $, the Schl\"afli symbol can be written in terms of dihedral angles as Eqs. (\ref{eq:p_da})--(\ref{eq:r_da}).
For a regular $D$-polytope $ \left\{ p_2 , \cdots , p_D \right\} $, if we introduce $\vartheta_0 = \vartheta_1 =0$ and denote by $\vartheta_i$ a dihedral angle of an $i$-dimensional face for $i \geq 2$, we can express the relation between Schl\"afli symbol and dihedral angles generally, as 
\begin{align}
\label{eq:pi_da}
p_i = \frac{\pi}{\arccos \left( \cos \frac{\vartheta_{i-1}}{2} \sin \frac{\vartheta_i}{2} \right) }.
\end{align}
As can be seen from Eq. (\ref{eq:pi_da}), the Schl\"afli symbol can be 
extended for $ i = 0, 1 $ as $p_0 = p_1 = 2$. Then, a regular $D$-polytope 
can be associated with a set of $D+1$ parameters 
$ \left\{ p_0 , p_1 , \cdots , p_D \right\} $.

Let us denote half the length of a line segment by $R_1$, and the radius of the circumsphere of a $D$-polytope by $R_D$.
Using an extended Schl\"afli symbol $ \left\{ p_0 , p_1 , \cdots, p_D \right\} $, we can write the first four of the circumradii as
\begin{align}
\label{eq:rad1}
R_1 &= \frac{l}{2} \sqrt{ \frac{ \sin^2 \frac{\pi}{p_1} }{ \sin^2 \frac{\pi}{p_0} } } , \\
\label{eq:rad2}
R_2 &= \frac{l}{2} \sqrt{ \frac{ \sin^2 \frac{\pi}{p_1} }{ \sin^2 \frac{\pi}{p_0} \sin^2 \frac{\pi}{p_2} } } , \\
\label{eq:rad3}
R_3 &= \frac{l}{2} \sqrt{ \frac{ \sin^2 \frac{\pi}{p_1} \sin^2 \frac{\pi}{p_3} }{ \sin^2 \frac{\pi}{p_0} \left( \sin^2 \frac{\pi}{p_2} - \cos^2 \frac{\pi}{p_3} \right) } } , \\
\label{eq:rad4}
R_4 &= \frac{l}{2} \sqrt{ \frac{ \sin^2 \frac{\pi}{p_1} \left( \sin^2 \frac{\pi}{p_3} - \cos^2 \frac{\pi}{p_4} \right) }{ \sin^2 \frac{\pi}{p_0} \left( \sin^2 \frac{\pi}{p_2} \sin^2 \frac{\pi}{p_4} - \cos^2 \frac{\pi}{p_3} \right) } } ,
\end{align}
where $l$ is the edge length of the regular polytopes.

From Eqs. (\ref{eq:rad1})--(\ref{eq:rad4}), the recurrence relations for the circumradii can be guessed.
In the expression of $R_i$, letting 
\begin{align}
\label{eq:rrcDp} 
\begin{cases}
\sin^2 \frac{\pi}{p_{i-1}} &\to~ \sin^2 \frac{\pi}{p_{i-1}} \sin^2 \frac{\pi}{p_{i+1}} \\
\cos^2 \frac{\pi}{p_{i-1}} &\to~ \cos^2 \frac{\pi}{p_{i-1}} \sin^2 \frac{\pi}{p_{i+1}} \\
\sin^2 \frac{\pi}{p_{i-2}} &\to~ \sin^2 \frac{\pi}{p_{i-2}} \left( 1 - \csc^2 \frac{\pi}{p_i} \cos^2 \frac{\pi}{p_{i+1}} \right) \\
\cos^2 \frac{\pi}{p_{i-2}} &\to~ \cos^2 \frac{\pi}{p_{i-2}} \left( 1 - \csc^2 \frac{\pi}{p_i} \cos^2 \frac{\pi}{p_{i+1}} \right)
\end{cases}
,
\end{align}
then we obtain the expression of $R_{i+1}$.
For the reader's reference we give the next three circumradii:
\begin{align}
\label{eq:rad5}
\textstyle R_5 &\textstyle= \frac{l}{2} \sqrt{ \frac{ \sin^2 \frac{\pi}{p_1} \left( \sin^2 \frac{\pi}{p_3} \sin^2 \frac{\pi}{p_5} - \cos^2 \frac{\pi}{p_4} \right) }{ \sin^2 \frac{\pi}{p_0} \left( \sin^2 \frac{\pi}{p_2} \left( \sin^2 \frac{\pi}{p_4} - \cos^2 \frac{\pi}{p_5} \right) - \cos^2 \frac{\pi}{p_3} \sin^2 \frac{\pi}{p_5} \right) } } , \\
\label{eq:rad6}
\textstyle 
R_6 &\textstyle= \frac{l}{2} \sqrt{ \frac{ \sin^2 \frac{\pi}{p_1} \left( \sin^2 \frac{\pi}{p_3} \left( \sin^2 \frac{\pi}{p_5} - \cos^2 \frac{\pi}{p_6} \right) - \cos^2 \frac{\pi}{p_4} \sin^2 \frac{\pi}{p_6} \right) }{ \sin^2 \frac{\pi}{p_0} \left( \sin^2 \frac{\pi}{p_2} \left( \sin^2 \frac{\pi}{p_4} \sin^2 \frac{\pi}{p_6} - \cos^2 \frac{\pi}{p_5} \right) - \cos^2 \frac{\pi}{p_3} \left( \sin^2 \frac{\pi}{p_5} - \cos^2 \frac{\pi}{p_6} \right) \right) } } , \\
\label{eq:rad7}
\textstyle
R_7 &\textstyle= \frac{l}{2} \sqrt{ \frac{ \sin^2 \frac{\pi}{p_1} \left( \sin^2 \frac{\pi}{p_3} \left( \sin^2 \frac{\pi}{p_5} \sin^2 \frac{\pi}{p_7} - \cos^2 \frac{\pi}{p_6} \right) - \cos^2 \frac{\pi}{p_4} \left( \sin^2 \frac{\pi}{p_6} - \cos^2 \frac{\pi}{p_7} \right) \right) }{ \sin^2 \frac{\pi}{p_0} \left( \sin^2 \frac{\pi}{p_2} \left( \sin^2 \frac{\pi}{p_4} \left( \sin^2 \frac{\pi}{p_6} - \cos^2 \frac{\pi}{p_7} \right) - \cos^2 \frac{\pi}{p_5} \sin^2 \frac{\pi}{p_7} \right) - \cos^2 \frac{\pi}{p_3} \left( \sin^2 \frac{\pi}{p_5} \sin^2 \frac{\pi}{p_7} - \cos^2 \frac{\pi}{p_6} \right) \right) } } .
\end{align}

\begin{table}[t]
  \centering
  \begin{tabular}{llll}\hline 
    & Regular simplex & Hypercube & Orthoplex \\ \hline
    $\left\{ p_2 \right\}$ & $\left\{ 3 \right\}$ & $\left\{ 4 \right\}$ & $\left\{ 4 \right\}$ \\
    $\left\{ p_2 , p_3 \right\}$ & $\left\{ 3 , 3 \right\}$ & $\left\{ 4 , 3 \right\}$ & $\left\{ 3 , 4 \right\}$ \\
    $\left\{ p_2 , p_3 , p_4 \right\}$ & $\left\{ 3 , 3 , 3 \right\}$ & $\left\{ 4 , 3 , 3 \right\}$ & $\left\{ 3 , 3 , 4 \right\}$ \\ 
    $\left\{ p_2 , p_3 , \cdots , p_{D-1} , p_D \right\}$ & $\left\{ 3 , 3 , \cdots , 3 , 3 \right\}$ & $\left\{ 4 , 3 , \cdots , 3 , 3 \right\}$ & $\left\{ 3 , 3 , \cdots , 3 , 4 \right\}$ \\ 
    $R_D$ & $\sqrt{ \frac{D}{ 2 \left( D+1 \right) } } l$ & $\frac{\sqrt{D}}{2} l$ & $\frac{\sqrt{2}}{2} l$ \\ \hline
  \end{tabular}
  \caption{The Schl\"afli symbols and the circumradii of regular simplices, hypercubes, and orthoplices. $l$ is an edge length of the polytopes.}
  \label{tab:hdrpt} 
\end{table}

It is well known that in five or higher dimensions there are only three types of regular polytopes: regular simplex, hypercube, and orthoplex.
Hereafter we restrict our investigation to these regular polytopes.
We summarize the Schl\"afli symbols and circumradii of the polytopes in Table \ref{tab:hdrpt}.
As can be seen from this table, the circumradii are written as simple functions in terms of dimension $D$.
Therefore the recurrence relations (\ref{eq:rrcDp}) can easily be inspected numerically.
In fact, we have put it into practice and confirmed the correctness for $D \leq 50$.

As can also be seen from Table \ref{tab:hdrpt}, $D$-simplex, $D$-cube, 
and $D$-orthoplex have the Schl\"afli symbol $ p_3 = \cdots = p_{D-1} = 3 $ in common. 
Thus in five or higher dimensions a general form of the circumradii of the $D$-polytopes might be given as a function of a set of three parameters $ \left\{ D , p_2 , p_D \right\} $.
Substituting $p_0 = p_1 = 2 $ and $ p_3 = \cdots = p_{D-1} = 3 $ into the functions generated by the recurrence relations (\ref{eq:rrcDp}) and the initial condition (\ref{eq:rad1}), and comparing the expressions, we can guess the general form of the circumradius of a regular $D$-polytope;  
\begin{align}
\label{eq:radD}
R_D = \frac{l}{2} \sqrt{ \frac{ \left( D - 2 \right) \cos \frac{2 \pi}{ p_D } - 1 }{ \cos \frac{2 \pi}{p_2}- \frac{D-3}{2} \cos \left( \frac{2 \pi}{p_2} -  \frac{ 2 \pi }{ p_D } \right) - \frac{D-3}{2} \cos \left( \frac{2\pi}{p_2} + \frac{2 \pi}{p_D} \right) + \cos \frac{2\pi}{p_D} } }
\qquad 
\left( D = 5,6,7,\cdots \right).
\end{align}
Assigning $ \left\{ p_2 , p_D \right\} = \left\{ 3 , 3 \right\} , \left\{ 4 , 3 \right\} , \left\{ 3 , 4 \right\} $ to Eq. (\ref{eq:radD}), the circumradii of $D$-simplex, $D$-cube, and $D$-orthoplex are reproduced, respectively.




\section*{Note added in proof}

At the end of this paper, we give a short account of the relation between Schl\"afli symbol and dihedral angles (\ref{eq:pi_da}).
We consider an arbitrary regular $n$-polytope $\left\{ p_0 , p_1 , p_2 , \cdots , p_n \right\}$
and denote it by $\Pi_n$ ,
where $p_0 $ and $ p_1$ are introduced in Appendix \ref{sec:crrdp}.
$ \Pi_n $ has regular $ (n-1) $-polytopes $ \left\{ p_0 , p_1 , \cdots , p_{n-1} \right\} $ as $ (n-1) $-dimensional faces.
Similarly each $ k $-dimensional face $\Pi_k$ has regular polytopes $\Pi_{k-1}$ for $ n \geq k \geq 1 $.

\begin{figure}[t]
  \centering
  \includegraphics[scale=1]{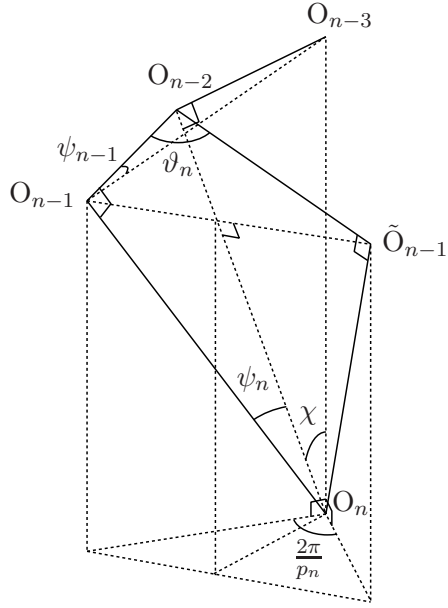}
  \caption{$\mathrm{O}_n$ is the circumcenter of $\Pi_n$.
  $\mathrm{O}_{n-1}$ and $\tilde{\mathrm{O}}_{n-1}$ are the centers of two $(n-1)$-dimensional faces sharing $\Pi_{n-2}$ centered at $\mathrm{O}_{n-2}$,
  and $\mathrm{O}_{n-3}$ is located at the center of $\Pi_{n-3}$.
  }
  \label{fig:dhatoss} 
\end{figure}

Let us choose a set of faces $ \Pi_0 , \Pi_1 , \cdots , \Pi_n $ satisfying $ \Pi_0 \subset \Pi_1 \subset \cdots \subset \Pi_n $ and denote by $\mathrm{O}_k$ the center of circumsphere of the $\Pi_k ~ ( n \geq k \geq 0 )$.
Furthermore we define the angles 
\begin{align}
\psi_n = \angle \mathrm{O}_{n-2} \mathrm{O}_n \mathrm{O}_{n-1}, \quad
\psi_{n-1} = \angle \mathrm{O}_{n-3} \mathrm{O}_{n-1} \mathrm{O}_{n-2}, \quad
\chi = \angle \mathrm{O}_{n-3} \mathrm{O}_n \mathrm{O}_{n-2}.
\end{align}
As can be seen from Fig. \ref{fig:dhatoss},
these satisfy 
\begin{align}
\tan \chi = \sin \psi_n \tan \psi_{n-1} , \quad
\tan \psi_n = \sin \chi \tan \frac{ \pi }{ p_n } , \quad
\vartheta_n = \pi - 2 \psi_n,
\end{align}
and from which we obtain Eq. (\ref{eq:pi_da}).




\end{document}